# Identifying the octupole Antiferromagnetic domain orientation in $Mn_3NiN$ by scanning Anomalous Nernst Effect microscopy

F. Johnson[1,a)], J. Kimák[2,a)], J. Zemen[3], Z. Šobáň[4], E. Schmoranzerová[2], J. Godinho[2,4], P. Němec[2], S. Beckert[5], H. Reichlová[4,5], D. Boldrin[1,6], J. Wunderlich[7] and L.F. Cohen[1]

1. Blackett Laboratory, Imperial College, Prince Consort Rd, London SW7 2AZ, United Kingdom

2. Faculty of Mathematics and Physics, Charles University, Prague, 121 16, Czech Republic

3. Faculty of Electrical Engineering, Czech Technical University, Technicka 2, Prague 166 27, Czech Republic

4. Institute of Physics, Czech Academy of Sciences, 18121, Prague, Czech Republic

5. Institut für Festkörper- und Materialphysik, Technische Universität Dresden, 01062 Dresden, Germany

6. SUPA, School of Physics and Astronomy, University of Glasgow, Glasgow G12 8QQ, United Kingdom

7. Institute of Experimental and Applied Physics, University of Regensburg, 93051 Regensburg, Germany

a) Corresponding authors: fj214@ic.ac.uk and jozef.kimak@mff.cuni.cz

Abstract

The intrinsic anomalous Nernst effect in a magnetic material is governed by the Berry curvature at the Fermi energy and can be realized in non-collinear antiferromagnets with vanishing magnetization. Thin films of (001)-oriented $Mn_3NiN$ have their chiral antiferromagnetic structure located in the (111) plane facilitating the anomalous Nernst effect unusually in two orthogonal in-plane directions. The sign of each component of the anomalous Nernst effect is determined by the local antiferromagnetic domain state. In this work, a temperature gradient is induced in a 50 nm thick $Mn_3NiN$ two micron-size Hall cross by a focused scanning laser beam, and the spatial distribution of the anomalous Nernst voltage is used to image and identify the octupole macrodomain arrangement. Although the focused laser beam width may span many individual domains, cooling from room temperature through the antiferromagnetic transition temperature in an in-plane magnetic field prepares the domain state producing a checkerboard pattern resulting from the convolution of contributions from each domain. These images together with atomistic and micromagnetic simulations suggest an average macrodomain of the order of 1 $\mu m^2$.

For spintronic applications, interest in AFMs has accelerated due to anticipated advantages in speed and energy efficiency they may offer over ferromagnets for memory and logic devices.[1,2] $Mn_3(A,B)N$ and related material families have been shown to support piezomagnetism [3], giant AHE [4] and MOKE [5] signatures plus unconventional torque geometries useful for out of plane magnetization switching.[6] Understanding and controlling the antiferromagnetic (AFM) domain structure is essential for application, although challenging to capture because of the compensated, near zero magnetization. By using a scanning laser to generate an out-of-plane thermal gradient, the anomalous Nernst effect (ANE) can be used as a powerful tool to image AFM domains/domain clusters, as previously demonstrated in $Mn_3Sn$ [7], and similarly domain structure has been imaged in the collinear AFM CuMnAs using the magneto-Seebeck effect.[8]

The ANE is one of a number of thermoelectric (TE) effects found in magnetic materials. The different geometries of the various TE effects are illustrated in Figure 1(a). Commercial TE devices exploit the conventional Seebeck effect, where a thermal gradient applied between two terminals generates a potential difference along the same direction.[9,10] In the Nernst effect (NE), the electric potential is perpendicular to both the applied temperature gradient and applied magnetic field [11], and is potentially attractive for commercial applications due to simpler thermopile design.[12,13] The anomalous Nernst effect manifests in ferromagnets, takes the same geometry as the NE, and is proportional to magnetization.[14] Originally thought to be present only in materials with finite magnetization, it is now accepted that there is an intrinsic ANE component associated with the Berry curvature of the band structure at the Fermi energy.[15] Experimental observation in non-collinear antiferromagnets (AFM) with vanishing magnetization, such as $Mn_3Sn$ [16-19] have helped cement this understanding. However, the magnitude of the intrinsic ANE in AFMs is found to be modest compared to conventional thermopower.[20,21] Improvement requires manipulation of both the Berry curvature and the Fermi level – unlike the anomalous Hall effect, maximizing the ANE requires topological features in the band structure to be moved away from the Fermi energy.[22,23,24] The family of nitride antiperovskites $Mn_3(A,B)N$ is chemically flexible on the A and B sites, facilitating tuning of electronic properties by chemical means.[25] Furthermore, it has recently been predicted that $Mn_3NiN$ presents a giant anomalous Nernst conductivity [26] and it has been shown the Berry curvature in $Mn_3NiN$ is also tunable using strain.[3,27] These directions are as yet embryonic.

Here we employ the scanning ANE technique to examine magnetic domain structure in non-collinear cubic antiperovskite $Mn_3NiN$ thin films. $Mn_3NiN$ films of thickness 50 nm used in this study were grown using pulsed laser deposition at 400°C on single crystal (001)-oriented STO substrates as described in detail in ref. [3]. The films are oriented, such that the [001] crystal direction (c-axis) points out-of-plane and 2 µm wide Hall bars patterned by electron-beam lithography are oriented along the [100] crystal direction. A typical device and the experimental geometry are shown in Figure 1(b,c) . The out-of-plane thermal gradient was generated by a continuous wave (CW) laser, power 10 mW, wavelength of λ = 800 nm, modulated by a mechanical chopper at frequency of 1.6 kHz. The beam was focused on the sample by an objective lens to a 1.8 µm diameter beam. The objective lens was placed onto a 3D piezo positioner, which allows for scanning of the laser spot across the Hall bar with a precision of 10 nm and a step-size of 400 nm. The thermo-voltage was measured using a lock-in amplifier. Simultaneously, the intensity of the reflected light was recorded in order to identify the position of the focused laser beam during the experiment. A schematic of the measurement set up is included in the supplementary information (SI) Figure S1. The data shown in the paper was taken at 50 K on the right hand cross C2. The supplementary information Figure S2 shows data taken at 100 K on the left hand cross C1, and in the bar region in Figure S3.

Atomistic, micromagnetic and COMSOL simulations are used to inform our experiments. The details of these calculations are also included in the SI.

Growth on an STO substrate imparts a small compressive bi-axial in-plane strain of the order of 0.1% [3] and the film considered here enters the non-collinear AFM magnetic phase, termed $\Gamma^{4g}$, below the magnetic ordering temperature $T_N$ = 230 K.

From the Mott relation between the anomalous Hall conductivity (AHC) and the anomalous Nernst conductivity (ANC) [28,29,30] it is anticipated that the symmetry of the two effects should be the same. The Γ$^{4g}$ phase can be described as a linear combination of three cluster octupole moments, and the magnetic symmetry group is the basis for the non-zero intrinsic anomalous Nernst effect in the material.[31] This phase has eight possible domain arrangements, corresponding to the Γ$^{4g}$ spin orientation being located in one of the eight {111} planes, as shown in Figure 2. Manipulating these domains by external magnetic field is made possible by growth on the STO substrate, where the in-plane strain leads to canting of the local Mn moments, creating a net weak uncompensated moment (red arrow in Figure 2) – with each one of the eight domains having the moment lie in one of the eight equivalent <112> directions. By cooling the sample in an applied in-plane magnetic field this weak net moment can be aligned to the field direction, thereby facilitating control and measurement of individual AFM macrodomains.

From the particular symmetries it follows that the intrinsic ANE is permitted in both in-plane crystal directions [100] and [010] when the out-of-plane thermal gradient, $\nabla T_{[001]}$, is applied to the Hall cross using the focused laser beam. Therefore, by scanning the laser over the device and measuring the sign of the resultant thermo-voltage in these two directions simultaneously, hence referred to as $V_{[100]}$ and $V_{[010]}$, we can identify the arrangement of different macrodomains.

Cooling in magnetic field $B \parallel [100]$ favors the four variants of Γ$^{4g}$ that have the in-plane component of the weak residual magnetic moment aligned to either [110] or [1$\bar{1}$0] – these are Figure 2 (a,b,c,d) shown in the blue box. This is because the net moments of these four variants all have the same projection to the applied field. Cooling with $B \parallel [\bar{1}00]$ favors Figure 2 (e,f,g,h) shown in the red box. These groups of variants will have identical sign of $V_{[010]}$. This is clearly observed in our measurement presented in Figure 3. The Hall cross device labelled C2 in Figure 1(c) is first located and identified using the optical reflected intensity (Figure 3(a,b)). After cooling from 300K to 50 K in negative (positive) 0.5 T external magnetic field applied in the [100] direction the region shows a saturated negative (positive) $V_{[010]}$ (Figure 3(c,d)). The signal remains unchanged even after removing the magnetic field. The voltage $V_{[010]}$ has the same geometry of the ANE as measured in a typical ferromagnetic system, and hence can be considered the "conventional" ANE.

From the signal shown in Figure 3 alone we cannot distinguish between the four possible variants of micromagnetic domains (Figure 2(a)-(d)). However, by measuring the "unconventional" ANE signal $V_{[100]}$ on the perpendicular set of contacts (Figure 4) further magnetic structure can be identified, with positive and negative regions ~1 µm in diameter arranged in a alternating pattern. This unconventional $V_{[100]}$ response corresponds to the blue and red highlighted components of the anomalous Nernst tensor in Figure 2. The form of the tensor was derived using the Linear Response Symmetry analysis tool and assuming the Mott formula. [32] The variants with [010] components to their moment (Figure 2(a,c,f,h)) produce a positive $V_{[100]}$ signal, while the variants that have [0$\bar{1}$0] components (Figure 2(b,d,e,g)) produce a negative signal. The sign of $V_{[010]}$ combined with the sign of $V_{[100]}$ therefore allows us to identify each region as corresponding to the two variants with identical in-plane component of the magnetic moment.

From symmetry considerations, it is expected that the magnitude of the unconventional intrinsic ANE and conventional intrinsic ANE components should be comparable, but experimentally we find the unconventional ANE is approximately six times smaller (see Figure 3 and 4, respectively). This can be understood by taking into consideration the likely size of the domains. Recent work on other AFM systems using XMLD-PEEM [33], NVD magnetometry [34] or Kerr microscopy [35] have identified individual domains considerably smaller than the feature size shown in Figure 4. The spatial resolution of the scanning ANE technique is limited by the lateral dimension of the thermally diffused spot on the sample, likely to be greater than the CW laser spot itself (1.8 µm in diameter).

The features we observe are due to the convolution of the laser beam as it scans across a mixed domain region, with individual domains that contribute positive and negative ANE voltage. This thus provides explanation for a reduced magnitude of the overall signal in the unconventional ANE direction compared to the conventional ANE measurement implying individual domains smaller than the laser spot size. However, were different domain types equally represented and significantly smaller than the spot size, this would result in a zero average ANE signal in the unconventional [100] direction. Clearly this is not the case, meaning that there are some overriding physical phenomena that encourages a propensity of domains of certain types. We consider four possible explanations: strain, antiferromagnetic domain walls, in-plane heat gradient and magnetostatics.

Strain from the constrained geometry of the Hall cross could in principle be responsible. But COMSOL calculations (see Supplementary for details) of the strain distribution indicate the strain due to lattice mismatch relaxation after patterning of the structure only extends 100 nm from the edges of the device. Therefore, it is likely to play a role only when devices an order of magnitude smaller than those considered here are made.

In $Mn_3Sn$, it has been suggested that the 180° antiferromagnetic domain wall (AFDW) is approximately 800 nm thick with a Bloch-like structure consisting of three 60° AFDW but the size and structure of the AFDW in $Mn_3NiN$ is unknown, and it is therefore plausible that the $V_{[100]}$ signal could be related to large domain walls. [36] Using atomistic simulations of strained $Mn_3NiN$, alongside prior density functional theory calculations (see Supplementary Information for further details), we simulate the 180° domain wall and parameterize it in terms of the strain induced weak ferromagnetic moment. [37] The parameters used in these simulations are summarized in Table 1. We also obtain a three-step domain wall, however with two 45° AFDWs occurring within the first and last steps, and a 90° AFDW within the central step. The width of each domain wall is ~8 nm, while the length of the full 180° AFDW is ~50 nm, significantly smaller than in $Mn_3Sn$, suggesting that the AFDWs do not provide a dominant contribution to the $V_{[100]}$ signal.

A further consideration to address for the finite voltage $V_{[100]}$ relates to the ordinary Seebeck due to the in-plane heat gradient $\nabla T_{IP}$, which is present at the edges of the laser spot and points radially outward from the center. When the laser is shone in the top left corner of the Hall cross C2, an in-plane heat gradient may be generated with components towards contacts T4 and T5. When the laser is shone in the bottom right corner, the in-plane heat gradient will point towards contacts T3 and T1. Therefore if the in-plane heat gradient was large and the ordinary Seebeck effect were visible, we would expect the top left corner and bottom right corner of C2 to present opposite sign of voltage in both the $V_{[010]}$ and $V_{[100]}$ scans, as the in-plane heat gradient has reversed sign. Yet, this is not what is observed, in fact both corners show the same sign of voltage. Furthermore, the ordinary Seebeck effect does not change sign in reversed field, yet the signal we observe shows a change in sign with field. Therefore, the signal cannot be attributed to ordinary Seebeck. (As shown previously with $Mn_3Sn$ [7], the domain structure may be rewritten if a region of the device is scanned using a higher laser power while a reversed field is applied, please see SI Figure S4 and associated text.)

Other effects that may occur in the presence of in-plane heat gradient are the magneto-Seebeck, and the component of anomalous Nernst associated with in-plane thermal gradients. The influence of these is discussed in detail in the supplementary. For better understanding of the macrodomain structure, we use also use COMSOL simulations to inform about the potential impact of this heat gradient. COMSOL calculations of heat propagation in the device (see Figure S5(a) in Supplementary) indicate this gradient is only significant for an extremely narrow region (of the order of 150 nm) compared to the out-of-plane heat gradient which is present over the entire laser spot. Its effect cancels when the laser spot is away from the edges of the device because the radially opposing directions of heat gradient generate opposite sign of ANE voltage.

Although the resultant octupole magnetization is modest, an order of magnitude smaller than typical ferromagnets, the resultant pattern that emerges from the ANE in the unconventional geometry is a checkerboard-type arrangement of domains constricted by the Hall cross, with area ~1 µm$^2$ indicative of a significant role of the magnetic dipole-dipole interaction (Figure 4(a, b). The extracted line scans from these images, which are shown in Figure 4 (c,d), reveal a gradual change from maxima to minima over the length of the device. Micromagnetic simulation of a film of 50nm thickness, with parameters taken from the atomistic simulation, yields a checkerboard-type arrangement of domains within the Hall cross geometry, with area ~1 µm$^2$ (Figure 4 (e)). If we then simulate the convolution process, taking the checkerboard pattern scanned with a stimulating laser field of 2 micron size, the resulting detected $V_{ANE}$ signal (Figure 4 (f)) strongly resembles the observed patterns (Figure 4 (a,b)) and with a reduced signal similarly of magnitude comparable to experiment. In contrast, $V_{[010]}$, the conventional component, is unaffected by the domain distribution because the contribution from all the different possible domain arrangements add constructively.

In conclusion, we have shown that the non-collinear Mn$_3$NiN films grown on STO substrates support the ANE, and that by using a scanning laser arrangement we can unveil the symmetries of the octupole macrodomain structure. We find conventional and unconventional components to the ANE due to the magnetic symmetry of the material. Although the method has proved ideal for spatial investigation of the domain arrangement, rigorous calibration of the temperature gradient would be required to determine the absolute value of the ANE. We note however, that the raw values are similar to those previously reported for Mn$_3$Sn films using the same experimental set-up. [7]. We finally determine that macrodomain clusters, present in the films, can be controlled through the magnetostatics associated with the modest net octupole moment.

See the Supplementary material for details of the simulations.

The data that support the findings of this study are available from the corresponding authors upon reasonable request.

We have no conflicts of interest to disclose.

Author contribution statement

F.J. and J.K. both contributed equally to this manuscript.

Acknowledgements

F.J acknowledges funding from Hitachi Cambridge and F.J and LFC from the UK Engineering and Physical Sciences Research Council (EPSRC). FJ and LFC acknowledge support from the Henry Royce Institute made through EPSRC grant EP/P02520X/1.

This work was supported in part by the Grant Agency of the Czech Republic under EXPRO grant no. 19 28375X and by EU FET Open RIA under grant no. 766566.

The work of J.Z. was supported by the Ministry of Education, Youth and Sports of the Czech Republic from the OP RDE program under the project International Mobility of Researchers MSCAIF at CTU No. CZ.02.2.69/0.0/0.0/18 070/0010457, and through the e-INFRA CZ (ID:90140).

D.B. is grateful for support from a Leverhulme Trust Early Career Fellowship (No. ECF-2019-351) and a University of Glasgow Lord Kelvin Adam Smith Fellowship.

| Parameter | Symbol | Value | Origin |
|---|---|---|---|
| Unit cell size | $a, b, c$ | $a = b$ = 3.882114 Å<br>$c$ = 3.895478 Å | [38], 0.1% strained values obtained with Poisson's ratio from [39]. |
| Exchange energy (NN) | $J_{ij}$ $Mn_1$ – $Mn_2, Mn_3$ | $-4.46 \times 10^{-21}$ J | *Ab. Initio* calculations to find unstrained values (see Supplementary) with strain adjustments inferred by fitting atomistic simulation of M(T) to experimental values of $T_N$ from [39]. |
| | $J_{ij}$ $Mn_2$–$Mn_3$ | $-5.16 \times 10^{-21}$ J | |
| Exchange energy (NNN) | $J_{ij}$ $Mn$ – $Mn$ in $a,b$ directions | $5.2 \times 10^{-22}$ J | |
| | $J_{ij}$ $Mn$ – $Mn$ in $c$ directions | $4.8 \times 10^{-22}$ | |
| Uniaxial anisotropy constant | $K$ – $Mn_1$ | $2.25 \times 10^{-23}$ J | [40], adjusted for strain. |
| | $K$ – $Mn_2, Mn_3$ | $2.55 \times 10^{-23}$ J | |
| Atomic spin moment | $\mu_s$ -$Mn_1$ | 2.83051 $\mu_B$ | *Ab. Initio* atomistic spin moments calculated for 0.1% strain. |
| | $\mu_s$ -$Mn_{2,3}$ | 2.82464 $\mu_B$ | |

S.B. and H.R. are supported by Czech Ministry of Education Grants LM2018110 and LNSM-LNSpin.

Table 1: Parameters used in atomistic simulations of strained $Mn_3NiN$, with their origins. $Mn_1$, $Mn_2$ and $Mn_3$ refer to the Mn spins located at (0.5,0.5,0), (0.5,0,0.5) and (0,0.5,0,5) respectively.

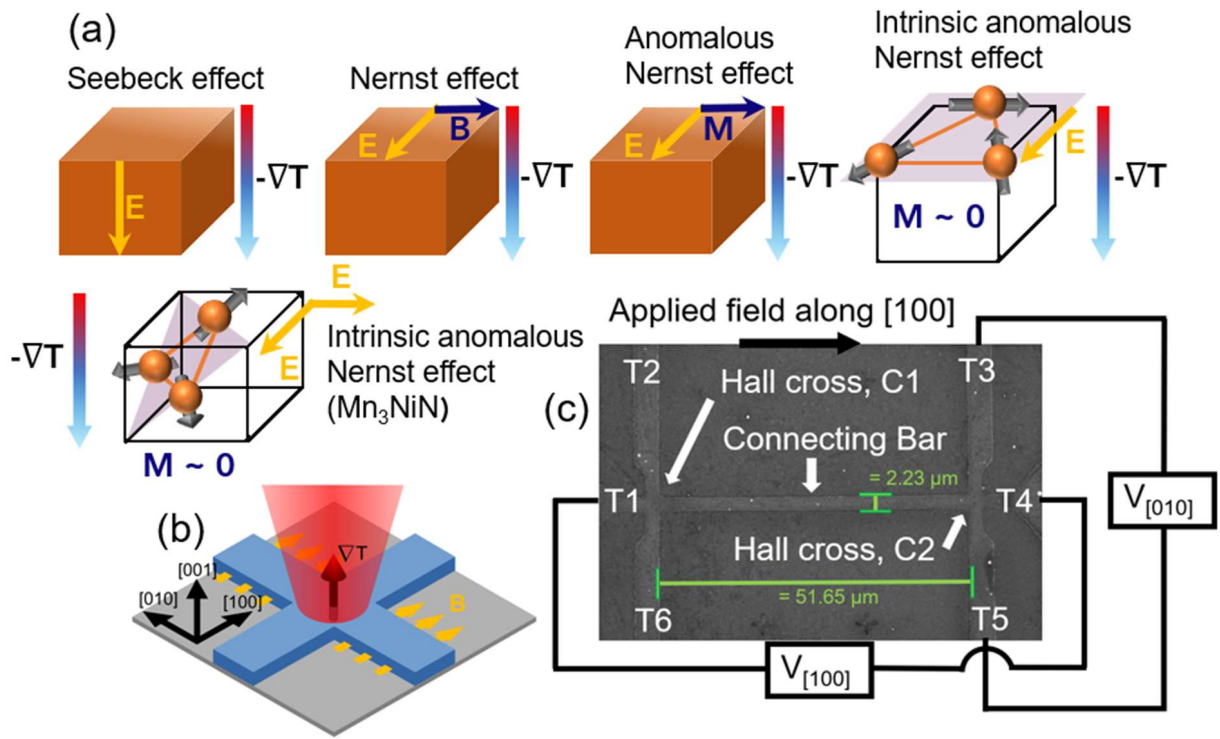

Figure 1: (a) Diagram of the various thermoelectric effects and their respective geometries. (b) Scematic of the scanning anomalous Nernst measurement of the $Mn_3NiN$ Hall cross. The laser is scanned across the surface generating a thermal gradient out-of-plane. The resultant thermovoltage in the two in-plane directions is simultaneously measured. (c) SEM image of device with overlaid dimensions, contacts T1-T6 and regions of the device labelled, and experimental geometry. $V_{[010]}$ is the conventional component of the anomalous Nernst effect, and $V_{[100]}$ is the unconventional component.

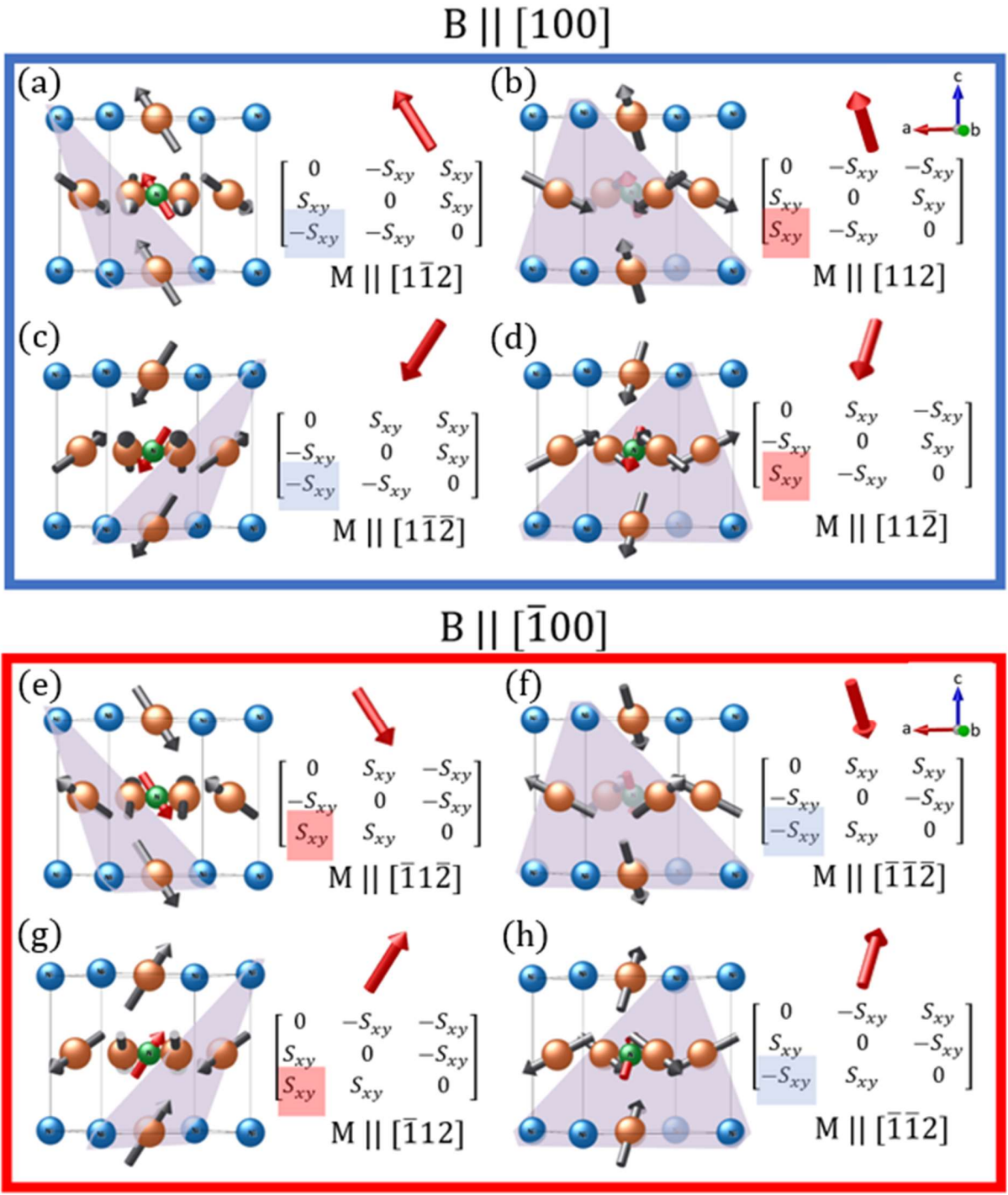

Figure 2: The 8 variants of Γ[4g], with the associated anomalous Nernst conductivity tensors and net moment directions. The first four variants in the blue box (a,b,c,d) are favored when B||[100] and give negative V[010] signal, but can give positive or negative V[100]. The other four in the red box (e,f,g,h) are favored for the case B||[$\bar{1}$00], and all have positive V[010] signal, but may also have either negative or positive V[100]. The coloring of the outer box shows the sign of the conventional Nernst component, while the highlighted component of the anomalous Nernst tensor shows the sign of the unconventional component. Drawings of the variants were produced using [41].

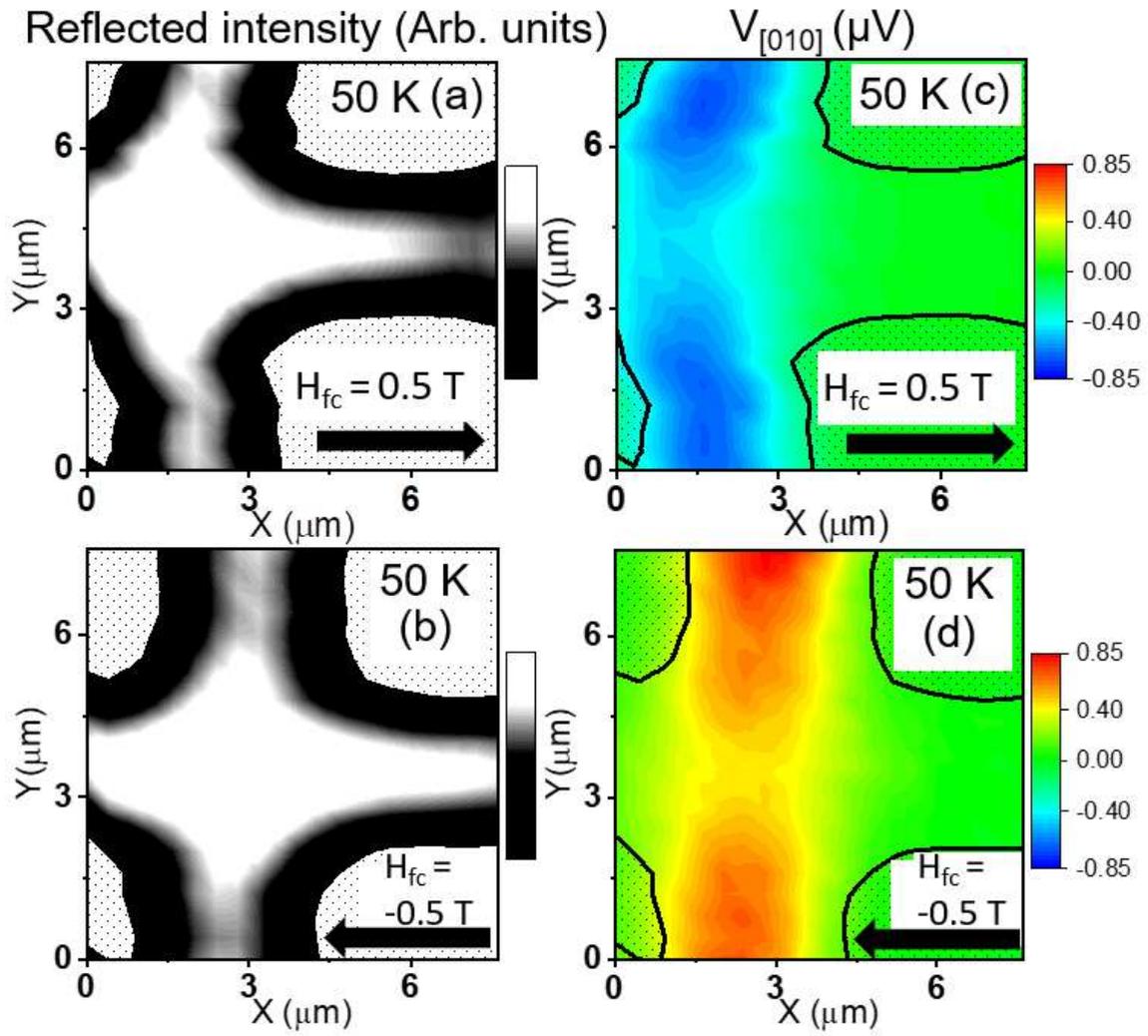

Figure 3: The conventional anomalous Nernst $V_{[010]}$ thermo-voltage in the Hall cross. (a,b) The reflected intensity of the area of the device scanned. (c,d) The $V_{[010]}$ thermo-voltage scans measured after cooling from 300 K to 50 K in positive and negative 0.5 T respectively, with the device area from the reflected intensity marked by black lines. In these scans, the region appears fully polarized.

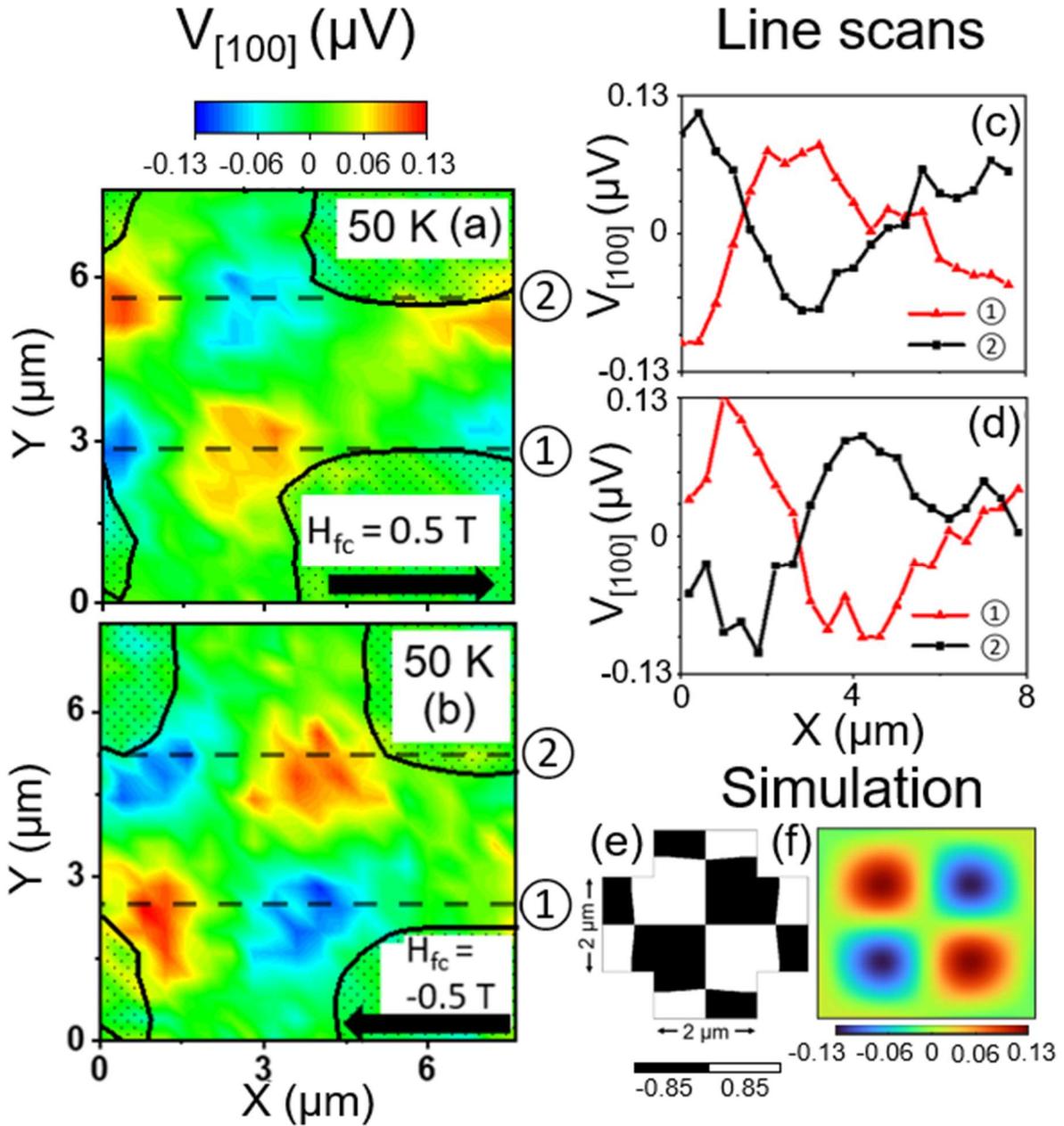

Figure 4: In the corresponding $V_{[100]}$ thermo-voltage scans (a,b) further magnetic contrast is visible, due to the underlying AFM domain clusters. The red and blue regions correspond with the positive and negative highlighted terms in the anomalous Nernst tensor in Figure 2. (c,d) Line scans taken from the $V_{[100]}$ thermo-voltage scans showing the width of the domain clusters. (e,f) Simulated $V_{[100]}$ scan for the checkerboard domain configuration with amplitude ± 0.85μV ((e) schematic) using Gaussian smoothing with a 1.8 μm FWHM to account for the focused laser spot. There is an overall amplitude reduction as well as a smoothing of the pattern.

Identifying the octupole Antiferromagnetic domain orientation in $Mn_3NiN$ by scanning Anomalous Nernst Effect microscopy

Supplementary material

Measurement schematic

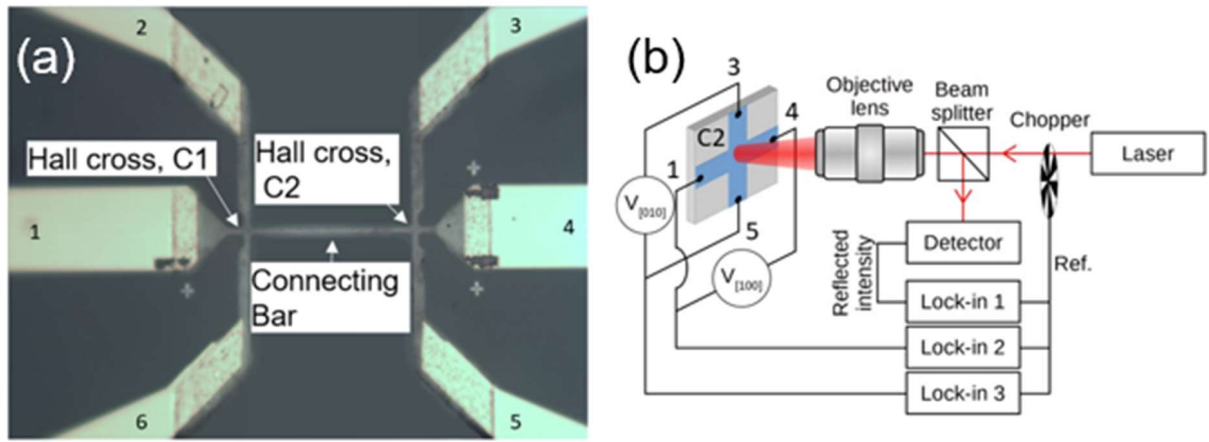

Figure S1: (a) SEM of the device structure showing the gold contacts (b) Schematic set up of the scanning ANE circuitry.

Note that figure S1 shows that measuring one of the active Hall crosses (either C1 or C2) could potentially introduce a measurable thermoelectric signal in $V_{[100]}$ from the asymmetry of the contacts. If it were an observable feature it would manifest as a subtle change in background due to the source of heat scanning across the sample, but certainly not in a checkerboard pattern and certainly not a switching pattern with field reversal. Hence we can rule this effect out as insignificant.

Data from left Hall cross C1 at 100 K

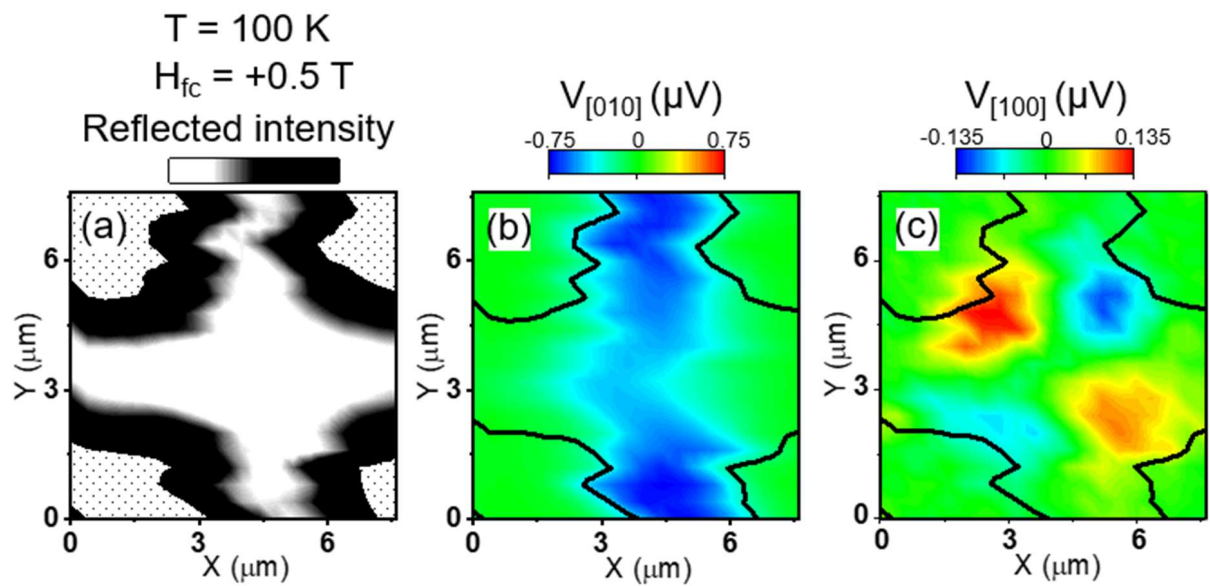

Figure S2 – Scans from the left Hall cross C2 at T = 100 K. Lower stability at this temperature has caused small distortions in the scans, but it is still clear the behavior of the left cross is the same as the right.

Data from the bar, and discussion of the in-plane heat gradient

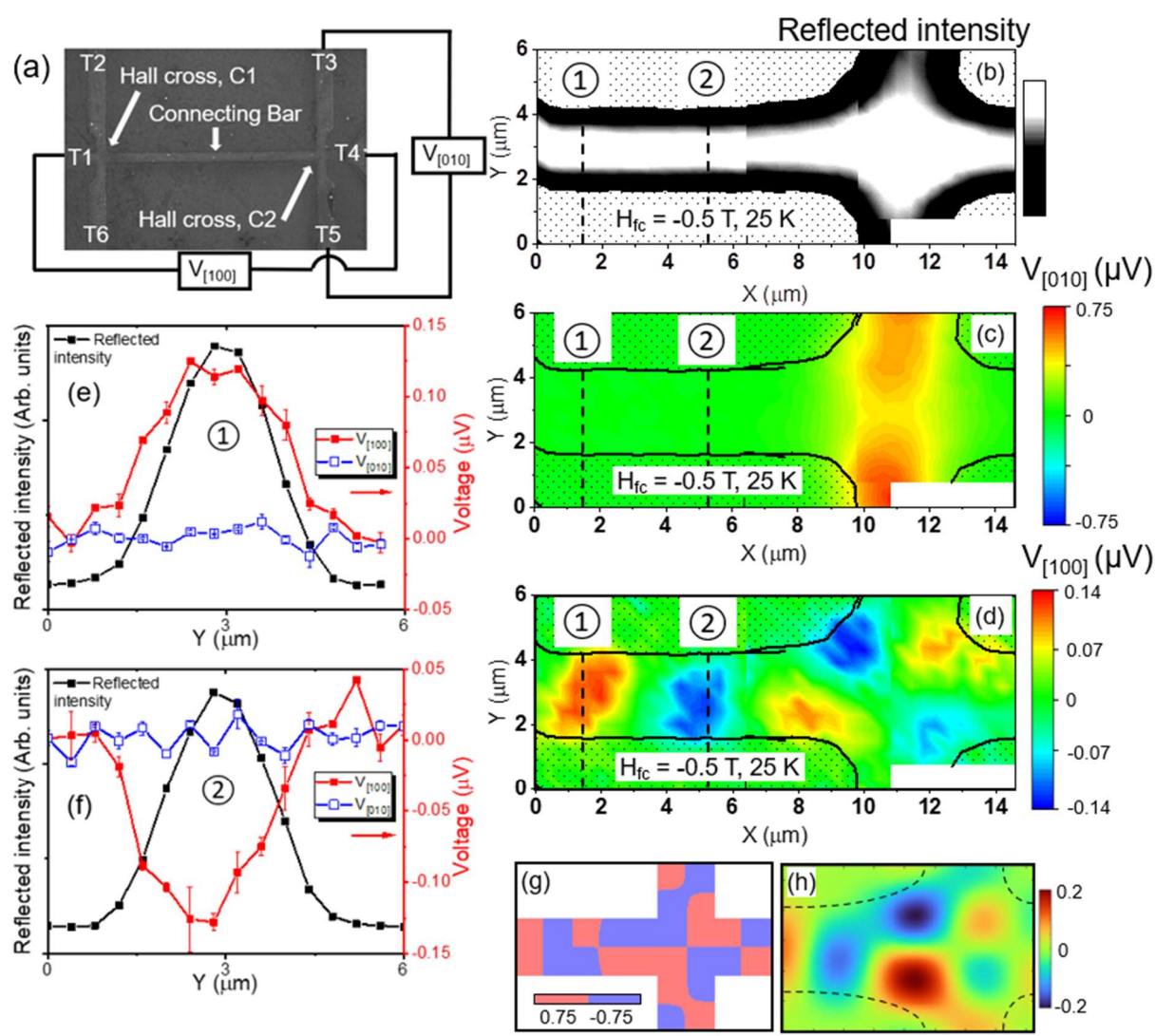

Figure S3 – ANE scans of the Hall cross and bar region of the device. (a) Image of the device with regions labelled. (b) Reflected intensity, showing the region of the scan where the bar is located. (c) $V_{[010]}$, measured using contacts 3 and 5, shows no signal in the region scanned as expected. (d) The $V_{[100]}$ scan along the bar also shows an alternating pattern of regions of positive and negative voltage. Images (b,c,d) are produced by combining three separate scans. (e,f) Line scan taken from the dashed lines on (b,c,d) show the maximal magnitude of voltage corresponds approximately with the center of the bar. (g) Micromagnetic simulations showing how domains in the device may be distributed. (h) Gaussian smoothing of (g) produces a pattern with similarities to what is observed in (d).

In-Plane Heat Gradients

A consideration to address for the finite voltage $V_{[100]}$ relates to the in-plane heat gradient $\nabla T_{IP}$, which is present at the edges of the laser spot and points radially outward from the center. There are two types of voltage signal we may detect that originate from in-plane heat gradient. The first is the component of the anomalous Nernst tensor that will create a voltage signal in the [100] direction when a heat gradient is applied along [010] (and vice versa). The second are Seebeck voltages – ordinary and magneto-Seebeck. We will address the component of ANE associated with in-plane heat gradient first.

In figure S3, we focus the laser along a section of the bar just outside the Hall cross. We firstly note that this region shows the unconventional voltage $V_{[100]}$ in an alternating pattern just like the Hall cross, but the conventional voltage cannot be measured, as there are no contacts in the [010] direction in this region. When the laser is shone along the bottom edge of the bar, the in-plane heat gradient will be directed along the positive Y direction (i.e. the [010] direction). In the center of the bar, the in-plane heat gradient points radially outwards along all directions, and therefore there will be equal in-plane heat gradient in all directions, and so the ANE from this in-plane heat gradient will cancel. And at the topmost edge, the in-plane heat gradient points in the negative Y direction (the $[0\bar{1}0]$ direction). Therefore, we expect the component of the ANE due to in-plane heat gradient to be maximal at the edges of the bar, to be positive at one end and negative at the other, and we expect to measure this effect in the $V_{[100]}$ direction. However, as can be seen from Figure S3 (d,e,f), maximal voltage is present at the center of the bar, not at the edges. Furthermore, no sign reversal of voltage is visible comparing the top edge and bottom edge. These experimental observations suggest the ANE due to in-plane heat gradient is small in comparison with the ANE due to out-of-plane heat gradient, and this agrees with the simulations that suggest the in-plane heat gradient is dwarfed by the out-of-plane heat gradient.

Now, we address voltage signals from the in-plane heat gradient via the Seebeck effect. In the bar measured in Figure S3 we do not expect to be able to measure the ordinary Seebeck effect, as the in-plane heat gradient is directed along the [010] along the bottom edge and $[0\bar{1}0]$ along the top edge, so voltage signals from this effect directed along these directions are not accessible. However, we anticipate the magneto-Seebeck effect to occur, as the magnetic order of the AFM structure breaks the transverse symmetry of the bar. This is exactly what was observed in CuMnAs, where a sample with the Néel vector at 45° to the bar direction showed opposite voltage along the bar boundaries due to the opposite net temperature gradient – please see Supplementary figure S5 of ref. [8]. Again, no such behavior is observed along the boundaries of the device, suggesting the observed voltage signals are dominated by effects from the out-of-plane thermal gradient.

Finally, we consider the influence of the ordinary Seebeck effect. As previously mentioned, this cannot be observed in the bar due to the geometry of the in-plane heat gradient with respect to the edges. The only place that we may be sensitive to it is in the corners of the Hall cross. As mentioned in the main text, the observed signals in the cross cannot be explained within the geometry of the ordinary Seebeck, and the ordinary Seebeck also does not change sign in reversed field. Therefore, the signal cannot be attributed to ordinary Seebeck.

To understand the observed domain pattern in the bar, we use OOMMF micromagnetic simulations [1]. The relaxed magnetic pattern is obtained assuming a single octupole moment representing each volume element with micromagnetic parameters derived from the atomistic simulation assuming three noncollinear moment in each unit cell. The result of the micromagnetic simulation is strongly dependent on the initial magnetization which can be set differently in each section of the device. In case of uniform initial magnetization along a direction in the range from [1 1 0] to [1 0.0001 0] the relaxed moments are all aligned with the easy axis [1 1 0] which gives negligible exchange energy and very low demagnetization energy. However, the single domain pattern is not observed experimentally. In case of uniform initial magnetization along [1 0 0] direction which is the hard axis as defined by the cubic anisotropy, the octupole moments are equally likely to relax towards the [1-10] and [110] easy axes which leads to the checkerboard pattern that is observed experimentally at least in the cross. This state is at a local energy minimum, which is much higher than the total energy of the single domain state with M along [110]. The relaxation in OOMMF at zero temperature cannot find the global energy minimum. In a real film, we expect the crystal axes in individual grains to be slightly misaligned so cooling in a field along the bar may effectively correspond to initial magnetization in individual grains along [1 0.0001 0] and [1 -0.0001 0] so the grains would relax to [1 1 0] and [1 -1 0], respectively - despite the higher cost in exchange and demagnetization energy.

In the bar to the left of the cross the single domain state with M along [110] is again more favorable than the continuation of the checkerboard pattern as expected, but its demagnetization energy is higher than that of a four-domain state after initialization in four sections with M along [110], [1-10], [110], and [1-10]. Assuming small misalignment of crystal axes in the grains of the film, we model a combined device of the bar and the cross with initial magnetization along [1 -0.0001 1] in one of the sections (with carefully selected location) of the bar and along [100] elsewhere. Such initialization in combination with minimization of the demagnetization energy returns a domain pattern which is consistent with the measured data. We do not claim that the simulation fully explains the measured domain structure. We show that a plausible distribution of crystalline anisotropy in the device combined with demagnetization effects can give rise to a magnetization pattern consistent with the experiment.

Heat assisted switching

A higher power laser may be used to demonstrate heat assisted switching of the AFM domains (Figure S4). The Hall cross C2 is cooled to 100 K in field $H_{fc}$ = 0.5 T, and the applied field direction is then reversed to $H_{app}$ = -0.5 T. As Figure S4 (b,c) shows, reversing the field at this temperature has no effect on the sign of the ANE, as the coercivity of the sample is too high in comparison with the small applied field. Then, the center of the cross is scanned with a 25 mW laser, raising the temperature sufficiently such that the coercivity drops and the area scanned is switched by the applied field. The area scanned in this way is indicated by the dashed box in Figure S4 (e,f). The new domain pattern is then measured using the laser at 10 mW in the usual method, and is shown in Figure S4 (e,f). In this way domain patterns can be written into the device.

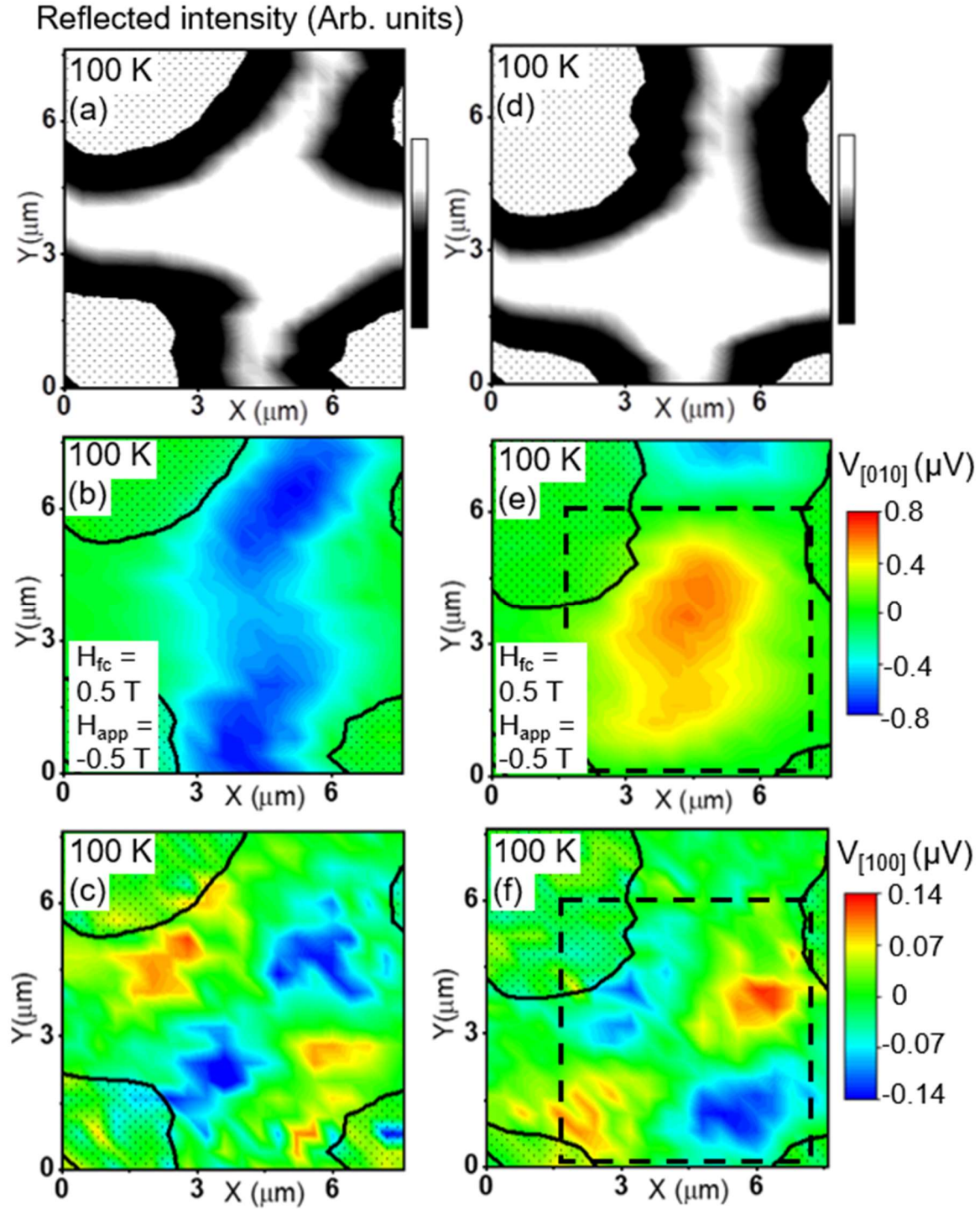

Figure S4: Heat assisted switching of the AFM domains. (a,b,c) – Scan of the Hall cross region C2 after cooling in $H_{fc}$ = 0.5 T and then reversing the field direction. The sign of the ANE shows no change when reversing the field at this temperature due to its large magnetic coercivity. (d,e,f) Scan of the same region after heating with increased laser power, showing a change of sign of ANE.

Atomistic simulation

An atomistic level model was used to describe the antiferromagnetic ordering of the $Mn_3NiN$ unit cell under 0.1% compressive strain in-plane, a = b = 3.882 Å, c = 3.895 Å, with Heisenberg exchange. We followed the approach used in the case of $Mn_3Ir$ [2,3,4] and described the system using the spin Hamiltonian,

$$H = \frac{-1}{2}\sum_{i \neq j} J_{ij} S_i \cdot S_j - K \sum_{\gamma} \sum_{k \subset \gamma} (S_k \cdot n_\gamma)^2 \quad (1)$$

Where $S_i$ are unit spin vectors at each site, $J_{ij}$ is the isotropic exchange coupling between the spins limited to nearest-neighbor and next-nearest-neighbor interactions. $K$ is the anisotropy constant and $\gamma$ represents the three Mn sublattices. $k$ is summed over all spins in sublattice $\gamma$, while $n_\gamma$ are unit vectors along the cubic axes directions. The parameters used are listed in Table 1 in the main text and are optimized by fitting the simulated $T_N$ to the experimentally found $T_N$ = 235 K for 0.1% strain in [39].

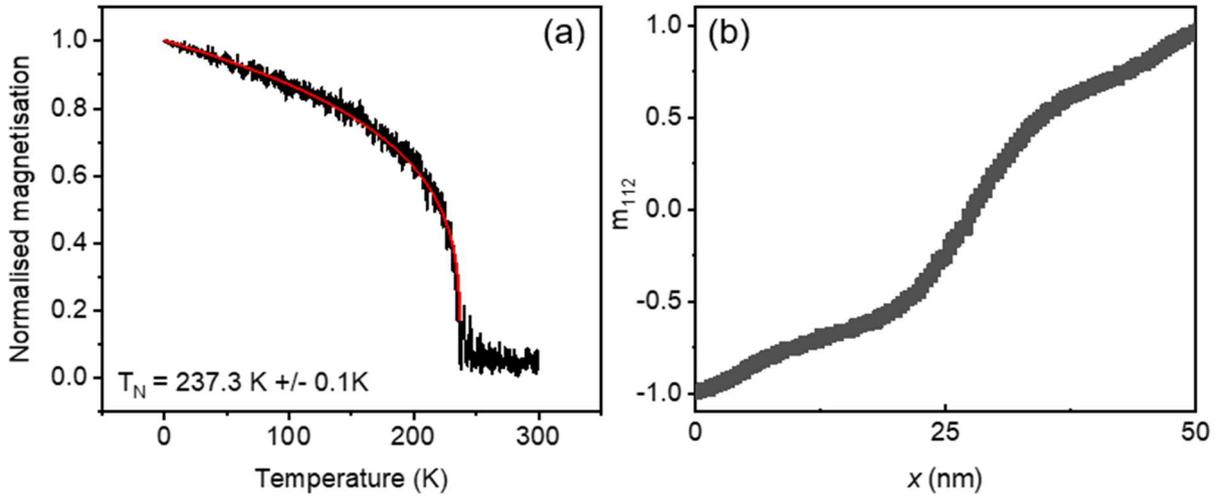

Figure S5: (a) Simulated M(T) using the parameters in Table 1. The resultant Néel temperature $T_N$ = 237.3 K is in good agreement with what is found experimentally in [38]. (b) Antiferromagnetic domain wall profile of a 180° domain wall.

To find the domain wall profile, we use the approach as detailed in references [5,6] and the constrained Monte Carlo method. [7] A system size with a cross section of 9x9 nm² and a length of 58 nm was used, with the top and bottom layers constrained to have net magnetization antiparallel, creating a Bloch-type domain wall at 0 K. All atomistic simulations were performed within the VAMPIRE software package. [8] By fitting the net magnetization domain-wall profile to the hyperbolic functions we extract the domain wall width $\delta_0$. At zero temperature, the energy of the domain wall, $\Delta E$, is simply the energy difference between the system with a domain wall and the system without. Then using the equations,

$$\delta = \pi \sqrt{\frac{A}{K}}, \quad (2)$$

$$\Delta E = 4\sqrt{AK}, \quad (3)$$

We obtained A = 9.68 × 10$^{-13}$ J/m, K = 5.97 × 10$^5$ J/m$^3$.

Micromagnetic simulation

Micromagnetic simulations in the paper were performed using MuMax3 [9] using cells of size 5 nm x 5 nm x 5 nm, saturation magnetization $M_s$ = 10000 A/m, cubic anisotropy K = -5.97 × 10$^5$ J/m$^3$ and exchange stiffness A = 9.68 × 10$^{-13}$ J/m. The system was initialized with uniform magnetization in the [100] direction, and the system was then relaxed to equilibrium using the adaptive time step RK23 solver implemented in MuMax3. As we expect only gradual rotations of the moments between domains and dynamics are not considered the net moment is a good description of each element of the micromagnetic simulation.

COMSOL simulation

We perform a simulation of heat transfer, thermal expansion and structural mechanics in our device utilizing the finite element method (FEM) as implemented in COMSOL Multiphysics®. [10] The software solves the coupled partial differential equations numerically using a tetrahedral mesh generated automatically for the 3D geometry of our device. The illumination by a laser is modelled as a uniform boundary heat source with circular shape of diameter = 1.5 µm active for 1 s (to approximate the laser scanning experiment). The peak power delivered to the spot is 5 mW which corresponds to the nominal experimental laser power of 10 mW combined with 50% heating efficiency (absorption). The boundaries of the substrate domain are set to a fixed temperature of 300 K, which acts as an efficient heat sink.

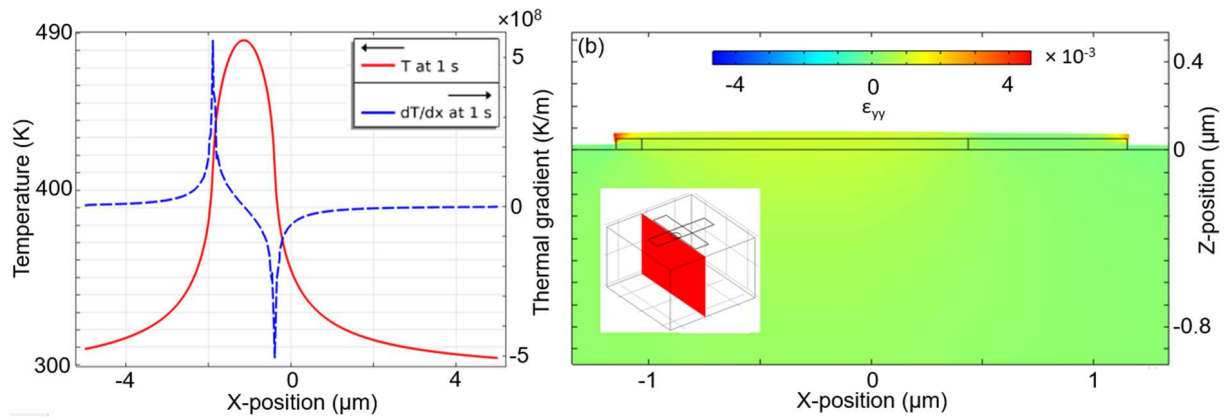

Figure S6: (a) The in-plane thermal gradient due to laser spot. The gradient is only significant in a region ~150 nm wide around the edge of the laser spot. (b) Strain, $\varepsilon_{yy}$, induced by patterning and by the laser spot for a y-z cut plane (see inset). The strain due to patterning only extends ~100 nm from the edges. (The deformation of the geometry is scaled up by factor of 10.)

| Parameter | Value | Origin |
|---|---|---|
| Thermal Conductivity | 12 W/m/K (STO) | [11] |
| | 3.2 W/m/K ($Mn_3NiN$) | [12] |
| Heat Capacity | 500 J/K/kg (STO) | [13] |
| | 600 J/K/kg ($Mn_3NiN$) | [14] |
| Thermal expansion | 9.4 ppm/K (STO) | [11] |
| | 20 ppm/K ($Mn_3NiN$) | [15] |
| Bulk modulus | 130 GPa ($Mn_3NiN$) | [16] |
| Youngs modulus | 70.2 GPa ($Mn_3NiN$) | [16] |
| Poisson's ratio | 0.41 ($Mn_3NiN$) | [17] |
| Elastic moduli | $C_{11}$ = 0.348 GPa, $C_{12}$ = 0.1 GPa, $C_{44}$ = 4.545 GPa (STO) | [18] |

Table S1: Material parameters used in COMSOL simulations.

DFT calculations

Calculations of the atomic spin moment and exchange energy employ the projector augmented-wave (PAW) method implemented in VASP code within the Perdew–Burke–Ernzerhof (PBE) generalized gradient approximation (no treatment of onsite electronic correlations was considered, the Hubbard U parameter was zero) – see methods of [15]. The canted triangular magnetic structure in bulk Mn3NiN subject to lattice strain was simulated.